\documentclass[pre,twocolumn,amsmath,amssymb,floatfix]{revtex4}

\usepackage{graphicx}
\usepackage{bm}
\begin{document}

\title{Atomistic simulation of structure and dynamics of columnar
  phases of hexabenzocoronene derivatives}

\author{Denis Andrienko}
\author{Valentina Marcon}
\author{Kurt Kremer}

\affiliation{Max-Planck-Institut f\"{u}r Polymerforschung,
  Ackermannweg 10, 55128 Mainz, Germany}

\date{\today}

\begin{abstract}
  Using atomistic molecular dynamics simulations we study solid
  and liquid crystalline columnar discotic phases formed by 
  alkyl-substituted hexabenzocoronene mesogens. Correlations between
  the molecular structure, packing, and dynamical properties of these
  materials are established.
\end{abstract}

\maketitle

\section{Introduction}

Most discotic thermotropic liquid crystals are formed by flat
molecules with a central aromatic core and several aliphatic chains
attached at the edges~\cite{ChandrasekharR90,BushbyL02}. The size and
 shape of the core can be varied, as well as the length and 
structure of the side chains, which helps to control the functional
properties of these materials~\cite{BrandKIM00}.
Triphenylenes~\cite{Kumar04}, hexabenzocoronenes (HBC) and their
derivatives~\cite{MullerKM98,BrandKIM00} are the most
frequently studied examples of discotics.

\begin{figure}
\begin{center}
\includegraphics[width=7cm]{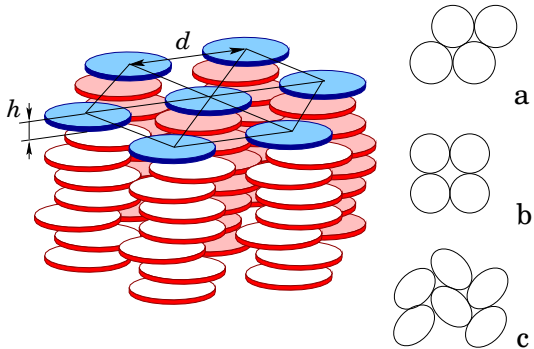}
\end{center}
\caption{ Columnar phase formed by the disc-shaped molecules and the most
  common arrangements of columns in two-dimensional lattices: (a)
  hexagonal, (b) rectangular, and (c) herringbone.}
\label{fig:columnar}
\end{figure}

The main feature of discotic liquid crystals is that they form
columnar phases, where the molecules stack on top of each other and
the columns arrange in a regular lattice~\cite{ChandrasekharR90} as
sketched in Fig.~\ref{fig:columnar}. The self-organization into stacks
with aromatic cores surrounded by saturated hydrocarbons results in
 predominantly one-dimensional charge transport within the core, along the
columns~\cite{vandeCraatsWFBHM99,Schmidt-MendeFMMFM01}. By modifying
the periphery and the shape of the core it is possible to obtain
materials with small charge trapping and recombination.

Conducting properties of discotics are useful for various photovoltaic
devices, in particular solar cells and plastic organic field-effect
transistors. In solar cells it is important that the charges can find
a direct percolation path to the electrodes without being trapped in
an island of the material. Therefore, the {\em morphology} of the
conducting film has a big impact on the efficiency of the
device~\cite{HallsAMWIWF00,ShaheenBSPFH01,AriasMSHIWRF01}. The ideal
structure is an aligned discotic liquid crystal with the columns
perpendicular (solar cell) or parallel (field-effect transistor) to
the substrate.

The spacial arrangement of stacks is, however, never perfect: the
columns can be misaligned, tilted, or form various types of
topological defects.  In addition, the local alignment of molecules in
columns can be different for different compounds, changing the overlap
of the $\pi$ orbitals and affecting the efficiency of the charge
transport in a single column. It is, therefore, important to control
both the local alignment of the molecules in the columns and the
global arrangement of the columns in the mesophase. Since these
properties are fixed by the molecular architecture and processing
conditions, the main task of synthetic chemistry is to synthesize
compounds which have various shapes of the core and different
peripheries, to optimize the conducting properties of the
mesophase. One should keep in mind, however, that, perfect 
arrangement of the molecules itself does not always mean optimal conducting properties: overlap of the $\pi-\pi$ orbitals, splitting of the HOMO level, etc. affect the efficiency of the charge transport and should be taken into account.

The actual synthesis of each particular compound can involve
many stages and be rather time-consuming. It is thus helpful to
get an insight into the mesophase properties prior to the actual
synthesis. For this purpose various computer simulation techniques can
be used. Depending on the length- and time-scales several techniques
can be involved: for example quantum chemistry calculations are
used to study the adsorption of molecules on electrodes and the charge
transfer mechanisms~\cite{BredasBCC04}; MD simulations with atomistic
force fields are employed to calculate local properties, such as
order parameters and molecular
arrangement~\cite{Maliniak92,OnoK92,MulderSPKdSK03,CinacchiCT04};
finally, coarse-grained simulations are successful in studying the
morphology of the bulk material, global arrangement of columns,
generic phase diagrams, and
defects~\cite{VeermanF92,EmersonLW94,BatesL96,Zewdie98,CinacchiT02,CaprionBR03,Bellier-CastellaCR04}.

The number of computer simulations so far performed on discotic
materials, is, however, relatively small and most of them are done
using idealized model potentials~\cite{BatesL96,CaprionBR03}. However,
a closer contact with real experiments requires more chemical details
to be put in. Up to now, there have only been a few attempts to perform
atomistic simulations of bulk discotic liquid crystals, mostly on the
triphenylene-based systems with a small aromatic
core~\cite{CinacchiCT04}.  Derivatives of hexabenzocoronenes are,
however, more promising for molecular applications, due to their
larger aromatic core and better overlap of the $\pi$-orbitals.  The
atomistic simulation of the structure and dynamics of the derivatives
of hexabenzocoronenes is the subject of the present paper.

\section{Studied systems and computational details}

\begin{figure}
\begin{center}
\includegraphics[width=8cm]{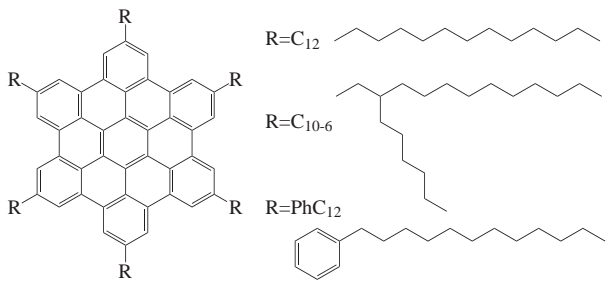}
\end{center}
\caption{Studied derivatives of hexabenzocoronene: R = $\rm
  C_{10}$, $\rm C_{12}$, $\rm C_{14}$, $\rm C_{16}$, $\rm C_{10-6}$,
  $\rm PhC_{12}$. Only R = $\rm C_{12}$ , $\rm C_{10-6}$ and $\rm
  PhC_{12}$ are shown.}
\label{fig:systems}
\end{figure}

The molecular structures of the studied hexabenzocoronene derivatives
are shown in Fig.~\ref{fig:systems}. All of them consist of a central
flat aromatic core and six side chains. We considered several types of 
side chains: alkyl chains of different lengths, ${\rm C}_n$, with
$n = 10, 12, 14, 16$; branched side chains, $\rm C_{10-6}$; and 
dodecylphenyl-substituted $\rm PhC_{12}$.  The choice of systems
is based on the availability of experimental data (wide angle
X-ray scattering, solid-state NMR, and differential scanning
calorimetry~\cite{FischbachPMFMSS02,vandeCraatsWFBHM99,FechtenkotterSHMS99,BrownSBMS99,HerwigKMS96}).

\subsection{Model potential}

\begin{figure}[ht]
\begin{center}
\includegraphics[width=8cm]{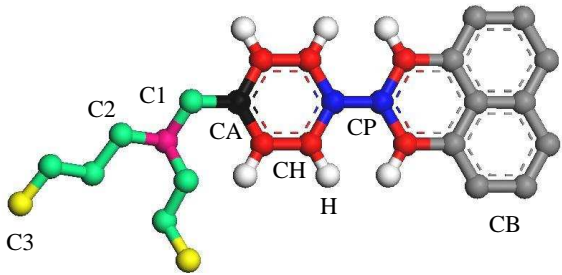}
\end{center}
\caption{Molecular fragment which illustrates the atom types used in the
    MD simulations. Methylene and methyl groups of the aliphatic chains are
    considered as one interaction site, according to the united atom
    approach.}
\label{fig:atoms}
\end{figure}

In our simulation we adopt the united atom approach, and consider
explicitly only the hydrogen atoms belonging to the aromatic rings of
the central core, since the charge on these hydrogens is responsible
for the multipole of the core of the molecule and the interaction of the
side chains with aromatic cores is not important in the columnar
phase. To describe the inter- and intra- molecular interactions, we
define eight different types of atoms (see Fig.~\ref{fig:atoms} and
Table~\ref{tab:nonbonded}): aromatic carbons of the central ring ($\rm
C_B$), aromatic carbons bonded to the hydrogens ($\rm C_H$), hydrogens
($\rm H$), aliphatic carbons representing either a methylene group
($\rm C_2$), terminal methyl group ($\rm C_3$) or a junction ($\rm
C_1$), and aromatic carbons bonding either alkyl chain ($\rm C_A$) or two
aromatic subsystems ($\rm C_P$).

The potential energy of the system is then given by the following sum
of contributions
\begin{eqnarray}
\label{eq:energy}
E &=&
\sum_{\rm bonds} \frac{1}{2} K_{b} \left( r - r_0 \right)^2 \\
\nonumber
&+& \sum_{\rm angles} \frac{1}{2} K_{\theta} \left( \theta - \theta_0 \right)^2 \\
\nonumber
&+&   \sum_{\rm dihedrals} \sum_{n=1 }^3 \left[ \frac{V_n}{2}
      \left(1+ (-1)^{n+1}\cos n \phi \right)
\right]  \\
\nonumber
&+& \sum_{\rm impropers } K_{d} \left( \psi -\psi_0 \right)^2  \\
\nonumber
&+&    \sum_{i}\sum_{j>i} \left[ \frac{1}{4 \pi \epsilon \epsilon_0} \frac{q_i q_j}{ r_{ij} } +
    4\epsilon_{ij}\left\{ \left(\frac{\sigma_{ij}}{r_{ij}}\right)^{12} -
     \left(\frac{\sigma_{ij}}{r_{ij}}\right)^{6}
\right\}
\nonumber
\right].
\end{eqnarray}
The first four terms in Eq.~\ref{eq:energy} correspond to the energy
due to the bonded interactions: harmonic bonds (1-2 interactions),
angles (1-3 interactions), and torsions, which include
Ryckaert-Bellemans and improper dihedrals (1-4 interactions). The
parameters of the bonded potentials are taken from the AMBER~\cite{Cornell95} and OPLS~\cite{Opls1,Opls2,Opls3} force fields and are listed in Tables~\ref{tab:bond}-\ref{tab:torsion}. 

The last double sum in Eq.~\ref{eq:energy} is due to the nonbonded
interactions. It is given by the Coulomb and Lennard-Jones terms with
$q_i$ the effective charges and $\epsilon_{ij}$ and $\sigma_{ij}$ the
well depth and contact distance of the sites $i$ and $j$. The
interaction potential between two atoms of the same type has its own
set of parameters $\{q, \epsilon, \sigma\}$, whereas between two different atoms the interaction energy is given by the geometrical mean of the two
pairs of atoms and the arithmetic average for the lengths $\sigma$ (Lorentz-Berthelot mixing rules).  Note that 1-2, 1-3, and also 1-4
interactions do not contribute to the above double sum. Furthermore,
nonbonded interactions between pairs of sites belonging to the central
aromatic core have been excluded. The parameters of the nonbonded
interactions are given in Table~\ref{tab:nonbonded}. Note that a
similar model has already been used to study
hexakis(pentyloxy)triphenylenes~\cite{CinacchiCT04}.

\begin{table}
\begin{tabular}{|c|c|c|c|c|}
\hline
 atom & mass (au) & $q (\rm e$) & $\epsilon (\rm kJ \cdot mol^{-1})$ & $\sigma (\rm nm)$ \\
\hline
 $\rm C_A, C_B, C_P$ & 12.011 & 0      & 0.293 &  0.355  \\
 $\rm C_H$ & 12.011 & -0.115 & 0.293 &  0.355  \\
 $\rm H  $ & 1.008  &  0.115 & 0.125 &  0.242  \\
 $\rm C_1$ & 13.019 & 0      & 0.450 &  0.3905 \\
 $\rm C_2$ & 14.027 & 0      & 0.494 &  0.3905 \\
 $\rm C_3$ & 15.035 & 0      & 0.732 &  0.3905 \\
\hline
\end{tabular}
\caption{Atom types and parameters of nonbonded interactions taken
from the OPLS force field. Atom types correspond to those shown in Fig.~\ref{fig:atoms}.}
\label{tab:nonbonded}.
\end{table}

\begin{table}
\begin{tabular}{|c|c|c|}
\hline
bond  &  $b_0 ({\rm nm})$ & ${\rm K}_b (\rm kJ \cdot mol^{-1} \cdot nm^{-2}) $ \\
\hline
 ${\rm C}_{ [\rm A,B,H,P] } - {\rm C}_{\rm [A,B,H,P] }$ & 0.139 & 418400 \\
 ${\rm C}_{ [2,3] } - {\rm C}_{ [2,3] }$         & 0.153 & 334720 \\
 ${\rm C}_{\rm H} - {\rm H}$             & 0.108 & 418400 \\
 ${\rm C}_{1} - {\rm C}_{2}$         & 0.153 & 334720 \\
\hline
\end{tabular}
\caption{Force constants and equilibrium bond distances. Atom types correspond to those shown in Fig.~\ref{fig:atoms}. Parentheses mean that no
  differences have been made between the included types of
  atoms.}
\label{tab:bond}
\end{table}

\begin{table}
\begin{tabular}{|c|c|c|}
\hline
angle  &  $\theta ({\rm deg})$ & ${\rm K}_{\theta} (\rm kJ \cdot mol^{-1} \cdot rad^{-2} )$ \\
\hline
 ${\rm C}_{[ \rm B,H]} - {\rm C}_{[\rm B,H,2]} - {\rm C}_{[\rm B,H,2]}$ & 120    & 527   \\
 ${\rm C}_{\rm 2} - {\rm C}_{\rm 2} - {\rm C}_{[2,3]}$         & 109.47 & 502.1 \\
 ${\rm C}_{\rm B} - {\rm C}_{\rm 2} - {\rm C}_{\rm 1}$       & 120    & 527   \\
 ${\rm C}_{\rm 1} - {\rm C}_{\rm 2} - {\rm C}_{2}$         & 109.47 & 502.1 \\
\hline
\end{tabular}
\caption{Force constants and equilibrium angles. Atom types correspond to those shown in Fig.~\ref{fig:atoms}. Parentheses mean that no
  differences have been made between the included types of
  atoms.}
\label{tab:angle}
\end{table}

\begin{table}
\begin{tabular}{|c|c|c|c|}
\hline
 dihedral & $V_1 $ & $V_2 $ & $V_3 $ \\
\hline
 ${\rm X} - {\rm C}_{[\rm A,B,P,H]} - {\rm C}_{[\rm A,B,P,H]} - {\rm X}$ & 0.0 & 121.34 & 0.0 \\
 ${\rm C}_{[\rm B, 2]} - {\rm C}_{[\rm 1,2]} -  {\rm C}_{\rm 2} - {\rm C}_{[2,3]}$  & 5.9 & -1.14 & 13.16\\
\hline
 improper & $K_d $ & $\psi_0$  &  \\
\hline
$\rm C_P - C_H - C_H - C_P$ & 167.4 & 0 & \\
$\rm C_H - C_H - C_A - H$ & 167.4 & 0 & \\
$\rm C_H - C_P - C_{[H,B]} - H$ & 167.4 & 0 & \\
$\rm C1 - C2 - C2 - C2$ & 334.8 & 35.26 & \\
\hline
\end{tabular}
\caption{ 
  Parameters of the torsional potential. X can be an aromatic carbon
  or a carbon bonded to a hydrogen.  Parentheses mean that no
  differences have been made between the included types of
  atoms. $V_1$, $V_2$, and $V_3$ are given in $\rm kJ \cdot mol^{-1}$
  for Ryckaert-Bellemans dihedrals and in $\rm kJ \cdot mol^{-1}
  \cdot rad^{-2} $ for improper dihedrals. The main task of improper dihedrals with $C_H$ atom type is to keep the corresponding   atoms in a plane. The $\mathrm{C1 - C2 - C2 - C2}$ improper constrains the tertiary carbon C1 to tetrahedrality. Atom types correspond to those shown in Fig.~\ref{fig:atoms}.}
\label{tab:torsion}
\end{table}

\subsection{Computational details}

We simulated systems of 160 molecules, arranged in columns of 10
molecules each. The initial configuration was either a hexagonal
($T=400\, \rm K$) or rectangular ($T=300\, \rm K$) arrangement of the
columns (see Fig~\ref{fig:columnar}). The distance between the columns
and between the molecules in the columns has been slightly increased
to avoid site superposition. After the energy minimization, a short
equilibrating run ($4\, \rm ns$) was performed at a constant pressure
$P = 0.1\, \rm MPa$ and a constant temperature $T=200\, \rm K$.
Quenching the system at a lower temperature helps to avoid trapping it
in a disordered (non-equilibrium) state from the very beginning.  We
used Berendsen thermostat~\cite{BerendsenPVDH84} with anisotropic
pressure coupling. The simulation box angles were fixed at $90\, \rm deg$. The electrostatic interactions were treated with the smooth particle-mesh Ewald method~\cite{essman1995} (spacing for PME grid $0.12\, \rm nm$, cutoff $0.9\, \rm nm$).

After the equilibration, the simulation box was a parallelepiped of
edges $L_x$, $L_y$, and $L_z$, with $L_x / L_y =
\sqrt{3}/2$ or $L_x / L_y = 1$, depending on the arrangement
of the columns (hexagonal or rectangular lattice).

The production run ($100\, \rm ns$) was performed at a constant pressure ($P = 0.1\, \rm MPa$) and temperature fixed using Berendsen thermostat with time constant $\tau_p = 5\, \rm ps$ and compressibility $4.5 \times 10^{-5}\, \rm Bar^{-1}$ (both isotropic). Two representative temperatures were chosen: $T = 400 \, \rm K$ and $T = 300\, \rm K$. The time step of $2\, \rm fs$ was used to integrate equations of motion using the velocity Verlet algorithm with a $0.9 \, \rm nm$ cut-off for short-range interactions. All calculations were performed with the parallel version of the GROMACS program~\cite{LindahlHv01}. Bond lengths were constrained using the LINCS algorithm~\cite{Hess97} provided by GROMACS.

\section{Results and discussion}

\begin{figure}
\begin{center}
\includegraphics[width=8cm]{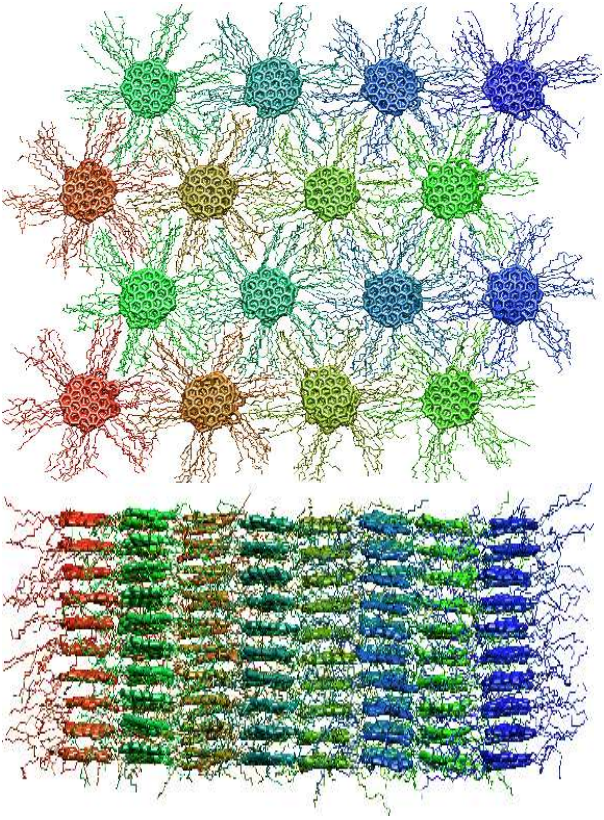}
\end{center}
\caption{MD simulation results: snapshot of the system with the $\rm
  C_{12}$ side chains. Aromatic cores are highlighted. Both top and
  side views are shown. $T = 400\, \rm K$, $P = 0.1\, \rm MPa$.}
\label{fig:snapshot}
\end{figure}

An equilibrated snapshot of one of the systems at $T=400\, \rm K$ is shown in
Fig.~\ref{fig:snapshot}. One can see that the aromatic cores are well
aligned and the columns are arranged in an almost perfect hexagonal
lattice, with the axially symmetric distribution of the side chains.
At $T=300\, \rm K$ the columns are arranged in a rectangular lattice and the
cores are also well aligned.  To further quantify the arrangement of
the molecules we calculated their orientational ordering as well as
several positional distribution functions.

\subsection{Orientational order}
To calculate the orientational order parameter of the columnar
mesophase we considered only the aromatic cores of the molecules. The
plane of the core was specified by the six vectors connecting the six
$\rm C_A$ atoms. The vector ${\bm u}^{(i)}$ was then defined as an
average of the six normals obtained by the cross product of every pair
of two neighboring $\rm C_A- C_A$ vectors. We calculated the
orientational order tensor for each column and for the whole system
\begin{equation}
Q_{\alpha \beta} = \left< \frac{1}{N} \sum_{i=1}^{N}
                   \left( \frac{3}{2} u^{(i)}_{\alpha} u^{(i)}_{\beta} -
                   \frac{1}{2} \delta_{\alpha \beta} \right) \right>,
\end{equation}
where ${\bm u}^{(i)}$ is a unit vector normal to the $i$-th aromatic
core, $N$ is the number of molecules in a single column or in the
whole system; correspondingly, the summation is performed either over
a single column or over the whole system. $<\dots>$ denotes the time
average. Diagonalizing this tensor we obtain the order parameter $Q$,
which is the largest eigenvalue of $Q_{\alpha \beta}$, and the average
orientation of the molecules, or director $\bm n$ (corresponding to
the eigenvector $Q$). $Q=1$ implies perfect alignment of molecules in
the column (all unit vectors parallel to each other), $Q=0$
corresponds to the isotropic angular distribution of the unit vectors.

\begin{figure}
\begin{center}
\includegraphics[width=8cm]{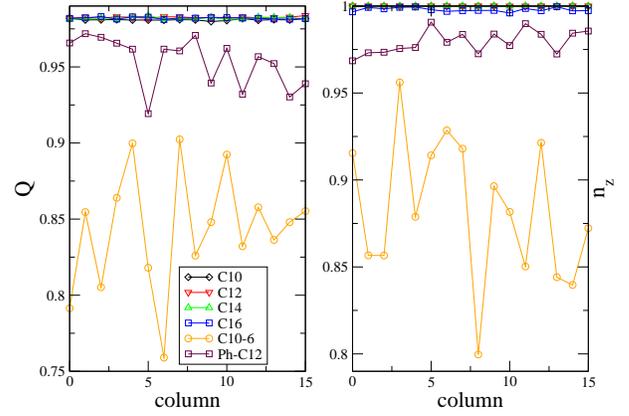}
\end{center}
\caption{Order parameter $Q$ and the $z$-component of the director
 (average tilt in the column) as a function of column number. $T=400\, \rm K$.}
\label{fig:order}
\end{figure}

The order parameter $Q$ and the average tilt angle of the unit vector
$\bm n$ (given by the projection onto the $z$-axis) are summarized in
Fig~\ref{fig:order}. It is clear that the systems with linear side
chains $C_{10}-C_{16}$ are well ordered: the order parameter is close
to its maximum value, $Q=1$; in addition, there is no tilt of the
molecules in the columns. The system with the
dodecylphenyl-substituted side chains, $\rm PhC_{12}$, has slightly
lower ordering (but still rather high, $Q \approx 0.95$) and a small
average tilt of the molecules in the columns. Finally, the system with
the branched chains, $\rm C_{10-6}$, has the worst ordering: the order
parameter is quite low and varies much from column to column;
there is also a significant tilt of the molecules in the columns.  As
can be seen in Table~\ref{tab:exp}, in the case of linear chains, 
the values of the order parameter $Q$ at $T=300\, \rm K$ are identical to the
one at $T=400\, \rm K$ , while the $\rm PhC_{12}$ shows a bigger disorder;
the $\rm C_{10-6}$ becomes more ordered with the decrease in temperature.

\subsection{Positional order}

To characterize the bulk structure of the columnar phase, we have
calculated two additional correlation functions. The first one, $g_z
(r)$, provides information about the arrangement of the molecules
in the planes parallel to the director $\bm n$, which in our case
is the direction of the columns (along the $z$ axis). In other words,
$g_z (r)$, is the probability to find a molecule at a distance $\bm n
\cdot \bm r$ from the reference molecule
\begin{eqnarray}
 g_{z}(r) &=& \frac{2}{V N}\sum_{\rm columns}\sum_{i=0}^{N_c} \sum_{j>i}^{N_c}
 f({\bm r}, {\bm n} \cdot {\bm r}_{ij}).
\end{eqnarray}
Here $N_c$ is the number of the molecules in a column, $r_{ij}$ is the
distance between the centers of mass of molecules $i$ and $j$, $N$ is the total number of molecules in the system. $f$ is a binning function, i.e. $f=1$ if ${\bm n} \cdot {\bm r}_{ij} \in [r \pm \Delta/2]$, where $\Delta$ is the bin size (in our case the number of bins was 400), otherwise $f=0$. $V = \Delta L_x L_y$ is the
volume of the bin. Note that the averaging is first performed in each column, and then over the columns, since the columns
can diffuse with respect to each other.

\begin{figure}
\begin{center}
\includegraphics[width=8cm]{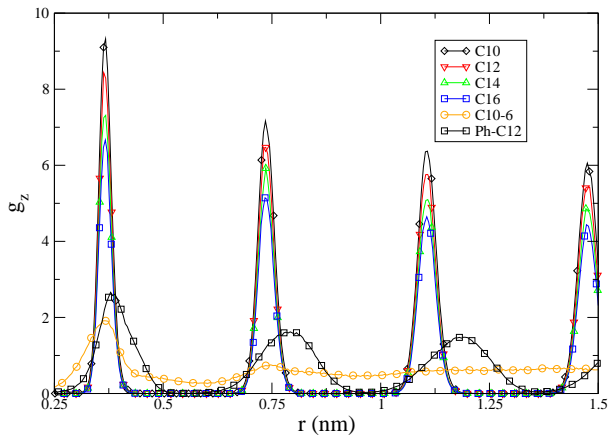}
\end{center}
\caption{Probability to find another molecule of the same column at a
   distance $r$ from the reference molecule. $T= 400\,\rm K$.}
\label{fig:rdfz}
\end{figure}

Figure~\ref{fig:rdfz} shows this distribution function for all studied
systems at $T=400\, \rm K$. It has a series of narrow peaks, which indicates
the ordered stacking of the molecules in the columns. For linear side
chains the peaks are sharp and well-defined, with a constant distance
between molecules in the columns. It is interesting that the 
separation is not affected by the length of the chain (of course,
in the range studied). If a phenyl ring is added, as it is in the
case of $\rm PhC_{12}$ or branched side chains, the peaks broaden and
shift, indicating a greater disorder of the mesophase.  For $\rm
PhC_{12}$ the distance between the columns also becomes slightly
bigger (see Table~\ref{tab:exp} for exact numbers).  Simulations at
$300\, \rm K$ show, for all the systems investigated, the same
distribution functions. This indicates that the difference of $100\,
\rm K$ in temperature is not affecting the ordering properties along
the columns. Intermolecular separations, which are given by the
distances between the neighboring peaks of the distribution function,
agree well with the X-ray scattering data and are summarized in
Table~\ref{tab:exp}.

The second distribution function, $g_{xy}$, provides information on
the arrangement of molecules in planes perpendicular to the director
$\bm n$, in the $xy$ plane
\begin{eqnarray}
 g_{xy}(r) &=& \frac{2}{V(r) N}\sum_{i=0}^{N} \sum_{j>i}^{N}
 f({\bm r}, {\bm n} \times  {\bm r}_{ij}),
\end{eqnarray}
where $V(r)=\pi \Delta (2R + \delta) L_z$ is the volume of a cylindrical bin.

\begin{figure}
\begin{center}
\includegraphics[width=8cm]{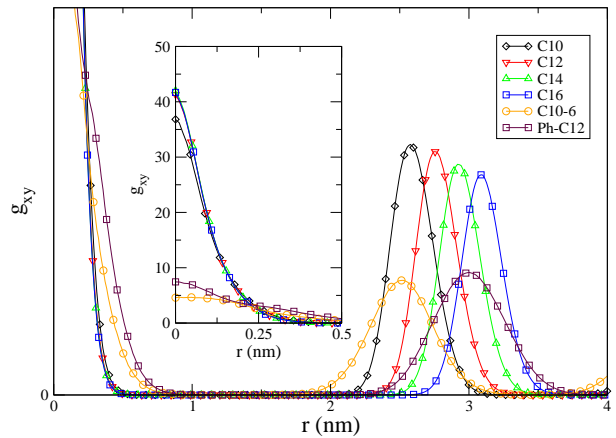}
\end{center}
\caption{Probability of finding the center of mass of another molecule at a
   distance $r$ from the reference molecule (two-dimensional radial
   distribution function). The inset shows the large peak around
   $r=0$. $T= 400\,\rm K$.}
\label{fig:rdfxy}
\end{figure}

The two-dimensional radial distribution function $g_{xy}$ is shown in
Fig.~\ref{fig:rdfxy}. It has a large peak around $r=0$, which is
because of the molecules in the same column. The other peak is due to
the in-plane nearest neighbors and provides us with the lattice
constant of hexagonal (or square) arrangement of the columns.

The spacing between in-plane (along the $z$-axis) nearest neighbors increases
 almost constantly with increasing length of the
linear side chains. In the case of branched chains, the position of
the peak is almost in the same position as the peak belonging to the
simulation of C10, but the peak is broader due to the disorder. For
$\rm PhC_{12}$ side chains the peak position is between C14 and
C16.

Both the measured values and the values estimated from X-ray diffraction experiments 
are presented in Table~\ref{tab:exp} and are in a fair agreement with
each other.

\subsection{Ordering of the side chains}

\begin{figure}
\begin{center}
\includegraphics[width=8cm]{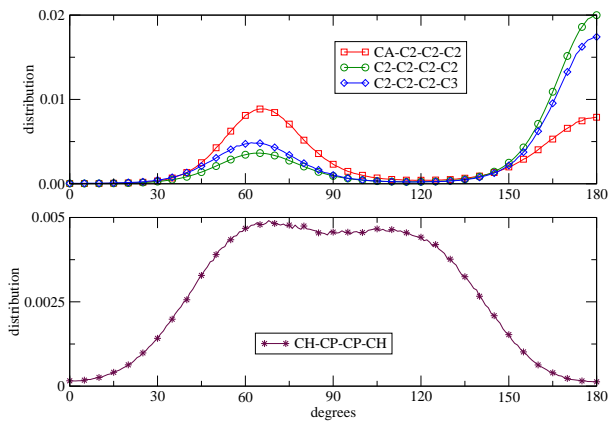}
\end{center}
\caption{Distribution function of the dihedral angles. The upper panel
refer to $\rm C_{12}$ run and the lower one to $\rm PhC_{12}$, both at T=400K.}
\label{fig:dih}
\end{figure}

To complete the description of the molecular arrangement in our systems, we also
 investigated the conformational state of the side chains and the 
degree of their interdigitation.
The conformational behavior of the alkyl side chains is shown in the upper
panel of Fig.~\ref{fig:dih}.
The distribution functions are given for the $\rm C_{12}$ system. The other systems show identical results.
The distribution function of the dihedral CA-C2-C2-C2 (see Fig.~\ref{fig:atoms} 
for the atom names), that is the dihedral connecting the core of the molecule with the alkyl side chains, shows that the gauche conformation (at around $60^{\circ}$) has the same probability as the trans one (at around $180^{\circ}$).
The distribution functions of the dihedrals along the alkyl chains (C2-C2-C2-C2) and
 the terminating one (C2-C2-C2-C3)show 
 that the trans state is more probable than the gauche one.
For the $\rm PhC_{12}$ system the distribution function of the CH-CP-CP-CH dihedral, shown in the bottom panel of Fig.~\ref{fig:dih}, indicates that the phenyl ring tends to align perpendicularly to the core of the molecule.

We have also evaluated the degree of interdigitation of the side chains. 
The results, which are shown in Table~\ref{tab:interdig}, were evaluated by subtracting the in-plane spacing between molecules in neighboring columns from twice the averaged end-to-end distances of the side chains and a (constant) distance between the center of mass of the molecule and the atom CA (which is approximately $0.57\, \mathrm{nm}$). The longer the side chain is, the higher the percentage of interdigitation is.

\begin{table}
\begin{tabular}{|c|c|c|c|}
\hline
system   & end-to-end (nm)  &  interdig. (nm) & interdig. (\%) \\
\hline
$\rm C_{10}$  & 1.02    & 0.61 & 59.8   \\
$\rm C_{12}$  & 1.19    & 0.76 & 63.8   \\
$\rm C_{14}$  & 1.35    & 0.91 & 67.4   \\
$\rm C_{16}$  & 1.51    & 1.07 & 70.9   \\
$\rm PhC_{12}$  & 1.50   & 1.14 & 76.0   \\
\hline
\end{tabular}
\caption{Interdigitation of the side chains.}
\label{tab:interdig}
\end{table}

\subsection{Translational dynamics}

To show that the systems have liquid-like dynamics, we 
calculated the mean square displacement of the molecular center of
mass, along the column direction ($z$ axis)
\begin{equation}
{\rm MSD}(t) = \left< \left[ z(t) - z(0) \right]^2 \right>,
\end{equation}
where the average is taken over all molecules.

\begin{figure}
\begin{center}
\includegraphics[width=8cm]{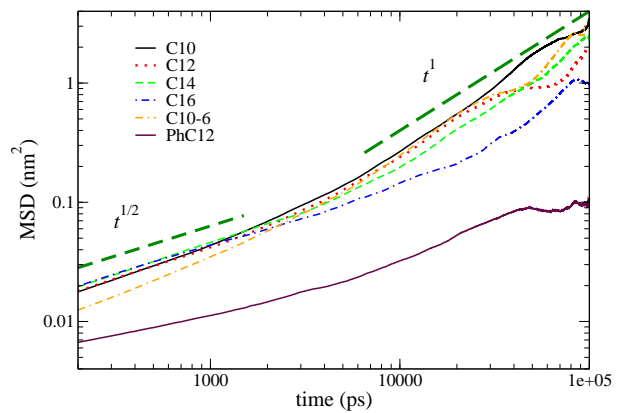}
\end{center}
\caption{log-log plot of the mean squared displacement. $T= 400\,\rm K$.}
\label{fig:msd}
\end{figure}

The resulting profiles are shown in Fig.~\ref{fig:msd}.  The log-log
profiles show that the mean-squared displacement has $t^{\frac{1}{2}}$
asymptotics for small $t$ which becomes $t^1$ in the long-time limit.
In fact, this type of asymptotics is expected for single-file
diffusion~\cite{richards:1977,kollmann:2003}, when individual
particles are unable to pass each other and the
particle order remains the same over time. In this case the motion of
individual particles requires the collective motion of many other
particles in the same direction.  In other words, in a one-dimensional
geometry the single-particle motion is coupled to the density
fluctuations. The fact that we can see the $t^1$ asymptotics of the
mean-squared displacement indicates that we already see diffusion
of the column as a whole. Note that, in an infinite one-dimensional
system there is no diffusion, since the diffusion coefficient scales
as $1/N$.

For $T = 300\, \rm K$ the log-log profiles show that the mean-squared displacement has different asymptotics (smaller than $t^{1/2}$) than for $T = 400\, \rm K$. However, the length of simulation runs does not allow us to conclude wether we have a frozen liquid crystalline state or a solid state. 

\section{Discussion and conclusions}

We first compare our results to the experimental data available in
Refs.~\cite{HerwigKMS96,FechtenkotterSHMS99,FischbachPMFMSS02}.

The X-ray diffraction experiments predict a columnar phase $D_h$
(hexagonal arrangement of columns) at $400\, \rm K$ and $D_r$
(rectangular arrangement) at $300\, \rm K$.  Our simulations show that
the systems at 400 K keep the hexagonal symmetry of the lattice during
the simulation run. The systems at 300 K stay rectangular for short
side chains; for C14, C16 systems the square lattice evolves into a
hexagonal one after about $2 \times 10^5 \rm ps$.

Concerning $xy$ spacing between the columns, the distance
increases with the length of the side chains. This trend seems not to
be present in experimental data, where, for example, the spacing for
C14 and C16 is identical. The deviation between the experimental data
and simulation results here is about 10\%.

The simulation data concerning the spacing between molecules in the
$z$-direction (along the columns) are in fair agreement with the
experimental results.  Maximum deviation for linear chains is about
3\%; for the phenyl case there is a deviation of about 10\%.

The order parameter obtained in simulations is systematically higher
than the one obtained experimentally. Of course, NMR data do not
provide exact numbers for the order parameter, rather an estimate of
it.  However, the main problem is the extremely small size of the
studied systems, since periodic boundaries do not allow
long-range fluctuations of the molecular order, as well as prohibiting the
presence of defects in the system.  Therefore, calculated values
of the order parameter merely emphasize the difference between
different systems, in a qualitative manner: the linear side chains
provide perfect ordering of the molecules in the columns; $\rm
Ph-C_{12}$ systems show good ordering; branched side chains introduce
a significant amount of disorder into the system, often leading to 
domains of material of different orientation.

Finally, translational dynamics shows typical single-file
diffusion behavior. It also indicates that one needs at least $10^5
\rm ps$ in order to calculate correct dynamic averages and to equilibrate
the systems.

The main message of this work is that atomistic molecular dynamics
simulations do not really provide complete quantitative description of the
columnar mesophases because of the very small time- and length-scales they can reach.
Hence, their main task is to serve as an input to coarse-graining techniques~\cite{AbramsK03,Muller-Plathe02,Izvekov2006,Voth_2006}, in
order to devise appropriate coarse-grained representations. One can then treat much bigger systems on longer timescales. This work is in progress.

\acknowledgments
We are grateful to Xin Zhou for the support at the early stages of this work. Advise of Benedict Reynolds is greatly appreciated.

\bibliographystyle{apsrev}
\bibliography{paper-cm}

\newpage 

\begin{table*}[ht]
\begin{ruledtabular}
\begin{tabular}{|c|c|c|c|c|c|c|c|c|}
          & $\rho$,400K       & $\rho$,300K        & d, 400K & d, 300K & h, 400K & h, 300K & Q, 400K & Q, 300K \\
          & $\rm g\, cm^{-3}$ & $\rm g\, cm^{-3}$  & nm      & nm      & nm      & nm      &         &       \\
\hline
C10-6     & \bf 1.001 & \bf 1.044 & \bf 2.54 & \bf 2.62   & \bf0.363  & \bf 0.331  & \bf 0.85  &\bf 0.97 \\
\hline
C10       & \bf 1.043 & \bf 1.087 & \bf 2.57 & \bf 2.26 & \bf 0.367 & \bf 0.363 & \bf 0.98 & \bf 0.98   \\
\hline
C12       &    &  & 2.86$\,$~\cite{FischbachPMFMSS02} & $2.38 \times 6.1$~\footnote[1]{this phase has the herringbone symmetry, see Fig.~\ref{fig:columnar}.}$^,$\cite{FischbachPMFMSS02} &
             0.355$\,$~\cite{HerwigKMS96} &  & 0.78\,~\cite{HerwigKMS96} &  \\
          &             &   & 2.52$\,$~\cite{HerwigKMS96}       &                                   &
             0.35$\,$~\cite{FischbachPMFMSS02}  & & 0.84$\,$~\cite{FechtenkotterSHMS99} & \\
          &  \bf 1.027  &  \bf 1.074 & \bf 2.76 &  \bf 2.53 & {\bf0.366} & \bf 0.365  & \bf 0.98 & \bf 0.98\\
\hline
C14       & &  & 2.66$\,$~\cite{HerwigKMS96}      & & 0.367$\,$~\cite{HerwigKMS96}       & &     &   \\
          &  \bf 1.012 & \bf 1.013 & \bf 2.93 & \bf 2.69  & {\bf 0.367 }  & \bf 0.362 & \bf 0.98 & \bf 0.98\\
\hline
C16       &  &  & 2.66$\,$~\cite{HerwigKMS96} &  &0.372$\,$~\cite{HerwigKMS96} & &    & \\
          & \bf 1.002 & \bf 1.002 & \bf 3.09  & \bf 2.87 & {\bf 0.368 }  & \bf 0.363 & \bf 0.98  & \bf 0.98 \\
\hline
Ph-C12    &  & & 3.4\,~\cite{FischbachPMFMSS02}  & 2.98\,~\cite{FischbachPMFMSS02} &
             0.35\,~\cite{FechtenkotterSHMS99}  & &0.93\,~\cite{FechtenkotterSHMS99}  & \\
          &          &            & &  &0.346\,~\cite{FischbachPMFMSS02}  &     &  &    \\
          & \bf 1.062  & \bf 1.103  &\bf 3.00 & \bf 2.69   &\bf 0.379  &  \bf 0.439  &\bf 0.95 &\bf 0.92
\end{tabular}
\caption{ 
   Summary of the experimental data and simulations. Results of the
   simulations are marked in bold.}
\label{tab:exp}
\end{ruledtabular}
\end{table*}

\end{document}